\documentclass[aps,prl,reprint,groupedaddress,longbibliography]{revtex4-1}
\usepackage{natbib}
\setcitestyle{super}
\usepackage{graphicx}
\usepackage{pgf}
\usepackage[english]{babel}
\usepackage{units}
\usepackage{physics}
\usepackage{float}
\usepackage{tikz}
\usetikzlibrary{shapes,arrows}
\usetikzlibrary{decorations.pathreplacing}
\usetikzlibrary{calc}

\begin{document}
\newcommand{\updown}{\uparrow \downarrow}
\newcommand{\upup}{\uparrow \uparrow}
\newcommand{\is}{\ket{F=1, m_F=0}}

\newcommand{\aw}{\mathrm{A}_\mathrm{W}}
\newcommand{\ar}{\mathrm{A}_\mathrm{R}}
\newcommand{\bw}{\mathrm{B}_\mathrm{W}}
\newcommand{\br}{\mathrm{B}_\mathrm{R}}

\renewcommand{\part}[1]{\paragraph{\textbf{#1}}}

\title{A network-ready random-access qubits memory}
\author{S. Langenfeld, O. Morin, M. K\"{o}rber, and G. Rempe}
\affiliation{Max-Planck-Institut f\"{u}r Quantenoptik, Hans-Kopfermann-Strasse 1, 85748 Garching, Germany}

\date{\today}
\begin{abstract}
Photonic qubits memories are essential ingredients of numerous quantum networking protocols. The ideal situation features quantum computing nodes that are efficiently connected to quantum communication channels via quantum interfaces. The nodes contain a set of long-lived matter qubits, the channels support the propagation of light qubits, and the interface couples light and matter qubits. Towards this vision, we here demonstrate a random-access multi-qubit write-read memory for photons using two rubidium atoms coupled to the same mode of an optical cavity, a setup which is known to feature quantum computing capabilities. We test the memory with more than ten independent photonic qubits, observe no noticeable cross talk, and find no need for re-initialization even after ten write-read attempts. The combined write-read efficiency is 26\% and the coherence time approaches 1ms. With these features, the node constitutes a promising building block for a quantum repeater and ultimately a quantum internet. 

\end{abstract}

\maketitle

\textbf{INTRODUCTION}\\

Quantum networks enable faithful communication by exchanging photonic qubits that cannot be cloned \cite{Wootters1982}. Networks conceived to generate an encryption key that is shared by two essentially classical parties, Alice and Bob, have already been demonstrated numerous times \cite{Zhang2018,Yin2020}. Future genuine quantum-mechanical and multi-functional networks, however, will likely rely on multi-purpose nodes that can receive, store, send and also process qubits \cite{Kimble2008,Wehner2018}. To accomplish these tasks, such nodes require at least two processable qubits \cite{VanEnk1997,Briegel1998,Childress2006,Jiang2007}.

For example, an elementary quantum repeater for long-distance entanglement distribution requires a chain of nodes, each with two memory qubits that are pairwise entangled with memory qubits in the neighbouring nodes. This entanglement is typically established via photonic channels which necessitate a corresponding qubit interface. Additionally, such repeater nodes require single- and two-qubit gates to also realize efficient entanglement swaps and complete Bell-state measurements. These properties, in essence, represent the celebrated five quantum computation plus two quantum communication criteria that were put forward by DiVincenzo \cite{DiVincenzo2000}. It needs to be emphasized that the criteria call for two contradictory physical capabilities \cite{Landauer1995}: memories need isolated qubits, while processors necessitate coupled qubits.

In order to realize these nodes, the qubits can either be distributed over an ensemble of (real or artificial) atoms or be localized in single (real or artificial) atoms. Important achievements for ensemble-based nodes include static-access (temporal) multiplexing where qubits are received and retrieved in a fixed order (e.g. first-in-first-out) \cite{Sinclair2014,Tang2015,Laplane2015,Tiranov2016}. This limitation can be resolved with spatial (dynamic-access) multiplexing instead of temporal multiplexing. In fact, a random-access quantum memory (RAQM) with an architecture resembling that of a classical random-access memory has been developed recently \cite{Jiang2019,Li2020}. This device can in principle receive and send an almost arbitrarily large number of qubits, limited only by the ratio of qubit coherence time to qubit access time. In the reported realization, this number was about three. However, no scalable information processing capability on and between the atomic qubits in all these ensemble nodes has been realized so far \cite{Saffman2016}.

Against this backdrop, it is still an open challenge to find a route towards multi-qubit memories that feature random qubit access, both for writing and reading, and controlled qubit processability. Individual real atoms are ideal to achieve this goal. They have already demonstrated, in complementary settings, their potential as faithful photonic qubit memories as well as qubit processors. The memory facet has been achieved by strongly coupling the atom to an optical cavity which enables the efficient interconversion between stationary (matter) and flying (photonic) qubits \cite{Reiserer2015}. Due to the high level of isolation from the environment and control of all degrees of freedom, these memories can feature long coherence times suitable for qubit storage \cite{Korber2018}. The processor facet has been demonstrated numerous times with qubits in atomic registers \cite{Wilk2010,Kaufman2015,Weiss2017,Levine2019,Bruzewicz2019} but without a connection to photonic qubits, or with a connection to photonic qubits \cite{Casabone2015,Inlek2017,Welte2018}, but in this case without multiplexed write-read capabilities.

Here we integrate the two facets by realizing an intracavity register with two individually addressable atomic memories that can independently write, store and retrieve photonic qubits. We demonstrate the atom-selective absorption and emission of these qubits by implementing several different random-access quantum-memory protocols with up to eleven photonic qubits that are subsequently stored in the node without the need for re-initialization. In addition to these new capabilities, the capacity of one qubit per single atom also sets a new benchmark on scalability for constant resources \cite{Nunn2008}.\\\\

\begin{figure*}[hbt]
\input{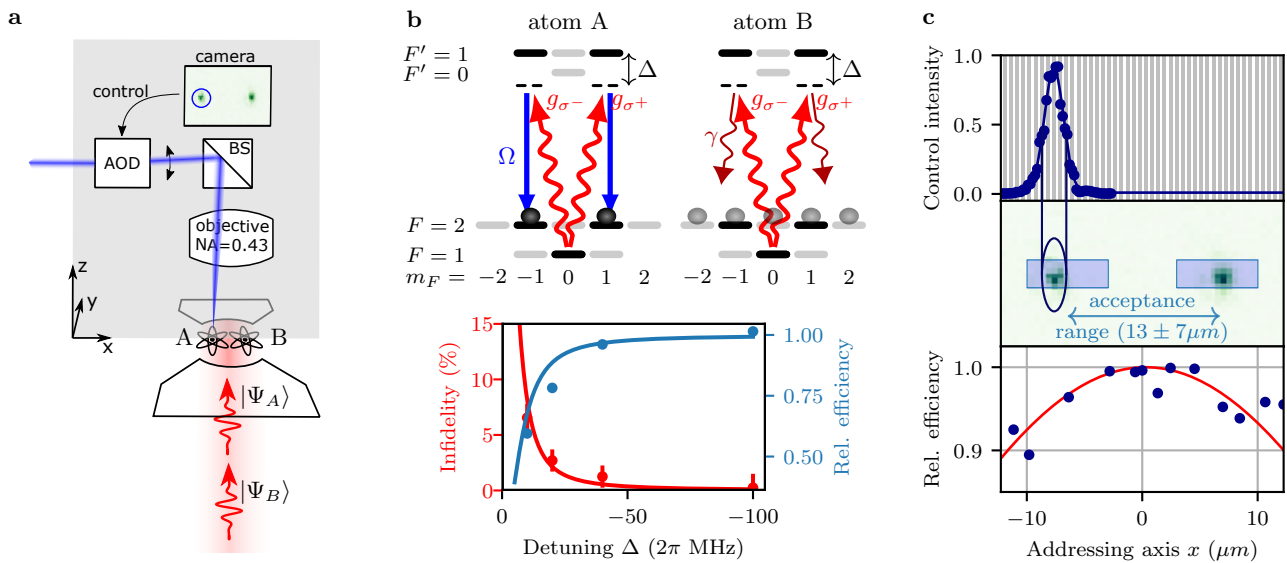}
\caption{\textbf{Illustrative view of the experimental setup and key elements towards the high-fidelity memory.}
\textbf{a}, 
Two atoms are trapped close to the waist of the cavity mode, symmetrically around the cavity axis.
Via a high-NA objective, the atoms are imaged on an EMCCD camera.
The same objective is used for addressing the individual atoms with a control laser beam by means of an acousto-optical deflector (AOD).
\textbf{b}, 
Possible cross talk between the atoms due to their coupling to the same cavity mode.
The state of an incoming photon should be mapped onto the addressed atom (A) via a stimulated Raman adiabatic passage (STIRAP) consisting of the cavity field $g$ and the classical control field $\Omega$.
However, the unaddressed atom (B) offers a side-channel via off-resonant scattering of the photon.
This leads to the storage of a random quantum state and thus a reduced fidelity.
The lower part of the figure shows experimental data and a theory curve for how this effect can be minimized by increasing the single photon detuning $\Delta$. The error bars correspond to one standard deviation of the statistical uncertainties. 
\textbf{c}, 
Cross-illumination of the addressing control field.
The control intensity as a function of the addressing axis is shown, including a Gaussian fit yielding a full width at half maximum (FWHM) of the addressing beam of $\unit[2]{\mu m}$.
As a reference, white and grey regions indicate the trapping potential pattern, in which the typical spatial extent of the atomic wavefunction is about $\unit[50]{nm}$.
For small cross-illumination, only atomic configurations are used where the inter-atomic distance is at least three times the FWHM.
The beam is elliptical (ratio 1:3) so that the atom positioning is less stringent in the direction of the long axis, illustrated by the ellipse in the second plot.
The spacing of the atom leads to a reduced atom-cavity coupling $g$ and thus to a reduced STIRAP efficiency (lower part).
The blue-shaded rectangles highlight the trapping regions used for this work.
}
\label{fig:setup}
\end{figure*}

\textbf{RESULTS}\\

\paragraph{Apparatus}\mbox{}\\

Figure \ref{fig:setup}a shows a sketch of the experimental setup. The system consists of two $^{87}$Rb atoms trapped close to the center of a high-finesse optical cavity with parameters $(g,\kappa,\gamma)/2\pi=\unit[(4.9,2.7,3.0)]{MHz}$. 
Here, $g$ denotes the single-qubit light-matter coupling constant at the center of the optical cavity for the $\is \leftrightarrow \ket{F'=1,m_F=\pm1}$ atomic transition at a wavelength of $\unit[780]{nm}$, and $\kappa$ and $\gamma$ are the cavity field and atomic dipole decay rates, respectively \cite{Korber2018}.

The two atoms are trapped in a two-dimensional optical lattice consisting of a red-detuned standing wave applied perpendicular to the cavity axis and a blue-detuned standing wave along the cavity axis. The two atoms are spaced symmetrically around the cavity center and therefore couple equally strong to the cavity mode. The two cavity mirrors have asymmetric transmission coefficients so that photons predominately leave the cavity via the high-transmission mirror. Outside the vacuum chamber, a high-numerical-aperture objective is used for imaging and laser-addressing of either atom \cite{Neuzner2016}.

Both atoms are initialized by optical pumping to the $\is$ ground state. 
Using a stimulated Raman adiabatic passage (STIRAP) technique \cite{Morin2019}, a single incoming photonic qubit is absorbed via the application of a classical control pulse. 
The information encoded in the photon polarisation (in the basis of right and left circular polarisation, $\sigma^+$ and $\sigma^-$, respectively) is coherently mapped to the two atomic states $\ket{F=2,m_F=1} $ and $\ket{F=2,m_F=-1}$, respectively. 
Thus, circular polarisations are mapped to atomic energy eigenstates whereas elliptical and especially linear polarisations are mapped to superpositions of atomic states. 
In order to reduce the impact of magnetic field fluctuations, a small magnetic guiding field of about $\unit[44]{mG}$ is applied along the cavity axis. 
Consequently, the phase of the atomic superposition oscillates at twice the Larmor frequency, $\unit[30]{kHz}$, during qubit storage, i.e., after write-in and before read-out. 
This oscillation is deterministic and could be compensated by a suitable setting of an optical wave plate. 
Finally, the stored photonic qubit is retrieved by vacuum STIRAP, the time-reversed process that is used for qubit storage. See supplementary material for more details on the timing of the experimental sequence.\\

\paragraph{Analysis and cross-talk elimination}\mbox{}\\

The main challenge for the realization of a high-fidelity multi-qubit quantum memory lies in avoiding cross talk between the individual qubit-memories. 
This means, first, that only the illuminated atom must couple to the incoming or the outgoing photonic qubit and, second, that cross-illumination of the atoms from the STIRAP control laser must be eliminated.
The first issue arises from the fact that both atoms are identically coupled to the same cavity field. 
Ideally, an incoming photon gets stored only in the addressed atom via the described STIRAP process (Fig. \ref{fig:setup}b). 
However, as all initialized atoms couple to the cavity, the incoming photon can interact with any atom including the unaddressed one. 
Scattering from this atom can transfer the atom into a random state and in this way scramble the fragile information encoded in the photon. 
A way out of this dilemma is that the STIRAP efficiency can be made constant over a large range of \textit{negative} single-photon detunings $\Delta$, \cite{Morin2019} and that the incoherent scattering rate scales like $\Delta^{-2}$. 
Quantitatively, we optimize the memory fidelity by increasing the detuning to $\Delta/2\pi=\unit[-100]{MHz}$ with respect to $\ket{F'=1}$, see Fig. \ref{fig:setup}b. 
At this point, the maximum incoherent scattering probability per input photon is only $\unit[(0.15\pm0.01)]{\%}$, in contrast to a coherent storage efficiency of $\unit[(38\pm3)]{\%}$.

For the second issue, cross-illumination is minimized by applying the STIRAP control field via an optical addressing system: 
After identifying the positions of the individual atoms via an electron multiplying charge coupled device (EMCCD) camera, the center of mass of the two atoms is shifted to the center of the cavity along one dimension (Fig. \ref{fig:setup}c). 
Note that the positions of the atoms are initially unknown. 
Then, a frequency-switchable radio frequency (RF) source is adaptively programmed so that each individual frequency drives the same acousto-optical deflector (AOD) in order to steer the beam onto one atom or the other (Fig. \ref{fig:setup}c). 
The pulse sequence on the atoms can be chosen randomly, with the only limitation being a delay time of $\unit[40]{\mu s}$ for switching between the atoms. 
This access time is mainly due to the sound-propagation time through the AOD, which could be reduced with, e.g., an electro-optical deflection system to about $\unit[2]{\mu s}$. 
All together this allows atom-number independent and thus scalable random access to the atoms. 
Practically, in order to reduce cross-illumination, we only use atom configurations where the inter-atomic distance is at least three times the full width at half maximum (FWHM) of the addressing beam along this direction, i.e. $d_\text{at-at} \geq \unit[6]{\mu m}$. For further details on the specific optical and electrical setup and on how the FWHM was measured at the atom's position, see Supplementary Information. 

A side effect of this separation is a small reduction of $g$ at the positions of the atoms. This results in a slightly reduced average memory efficiency of $(\unit[26\pm3)]{\%}$ for the combined storage and retrieval process (Fig. \ref{fig:setup}c). Nevertheless, the atom-to-photon state-transfer probability that is relevant, e.g., for atom-photon entanglement still exceeds $\unit[60]{\%}$ in our system. This is to the best of our knowledge larger than in any other multi-qubit memory that has been demonstrated so far.\\

\paragraph{Random-access memory with up to 11 qubits}\mbox{}\\

We probe the memory coherence and efficiency by storing weak coherent pulses (mean photon number $\bar{n}=1$) of different polarisations and performing a polarisation tomography on the retrieved single photon. 
In order to test for cross talk between the atoms, we probe atom $A$ and atom $B$ with identical and orthogonal polarisations. 
Table \ref{table_fidelities} summarizes the measured fidelities of the two polarisation configurations for both atoms in two different access patterns. 
Here, $A_W$, $B_W$ and $A_R$, $B_R$ stand for writing and reading, respectively, of one of the atoms, $A$ or $B$. 
In both access patterns shown in the table, storage of a photon in atom $B$ occurs when the qubit already stored in atom $A$ has rephased to its initial state. 
This timing is chosen to test for maximum cross talk that would manifest when the two input polarisations change from identical to orthogonal.

For the sequence $A_WB_WA_RB_R$, the minimum storage time for both qubits is $\unit[100]{\mu s}$, given by the Larmor frequency and the delay time of the addressing system.
The storage time for the data presented in Table \ref{table_fidelities} is $\unit[133]{\mu s}$. 
For both atoms we achieve high fidelities for eigen- and superposition states as input, with no notable difference compared to the single-atom case \cite{Korber2018}.

\begin{table}
\caption{Fidelities in percent of qubit memory implementations for 4 different input polarisation combinations and 2 random access possibilities. The errors correspond to one standard deviation of the statistical uncertainties.  
\label{table_fidelities}}

\def\element(#1,#2){
\begin{scope}[scale=2,shift={(2em*#2,0)}]
	\node[] at (0.5em,0.5em) {#1};
	\draw (-0.5em,0) -- (0,0) -- (0,1em) -- (1em,1em) -- (1em,0) -- (1em+0.5em,0);
\end{scope}
}

\def\elementS(#1,#2){
\begin{scope}[scale=2,shift={(2em*#2,0)}]
	\node[] at (0.5em,0.5em) {#1};
	\draw (0,0) -- (0,1em) -- (1em,1em) -- (1em,0);
\end{scope}
}

\def\wait2(#1){
	\draw ({-1em+4em*(#1)},0) -- ({3em+4em*(#1)},0);
}

\def\waitx(#1,#2){
	\foreach \pos in {0,...,#2}{
		\wait2(\pos+#1)
	}
}

\begin{tabular}{cccccc}
&\multicolumn{2}{c}{$\aw\bw\ar\br$} &\multicolumn{2}{c}{$\aw\bw\br\ar$}\\
\hline
atom A&\multicolumn{2}{c}{\begin{tikzpicture}[scale=0.6]
	\element(W,1)
	\waitx(2,0)
	\element(R,3)
	\waitx(4,0)
\end{tikzpicture}}
&
\multicolumn{2}{c}{\begin{tikzpicture}[scale=0.6]
	\element(W,1)
	\waitx(2,0)
	\element(R,4)
	\waitx(3,0)
\end{tikzpicture}}
\\
atom B&\multicolumn{2}{c}{\begin{tikzpicture}[scale=0.6]
	\element(W,2)
	\waitx(1,0)
	\waitx(3,0)
	\element(R,4)
\end{tikzpicture}}
&\multicolumn{2}{c}{\begin{tikzpicture}[scale=0.6]
	\waitx(1,0)
	\element(W,2)
	\element(R,3)
	\waitx(4,0)
\end{tikzpicture}}
\\
\hline
input pol.& atom A & atom B & atom A & atom B\\
\hline
$\ket{R}_\text{A}, \ket{R}_\text{B}$ & $ 97.0\pm0.7 $ & $ 96.8\pm0.6 $ & $ 97.2\pm0.4 $ & $ 97.3\pm0.3 $\\
$\ket{R}_\text{A}, \ket{L}_\text{B}$ & $ 97.8\pm0.6 $ & $ 96.4\pm0.4 $ & $ 97.0\pm0.3 $ & $ 98.1\pm0.5 $\\

$\ket{H}_\text{A}, \ket{H}_\text{B}$ & $ 95.4\pm1.3 $ & $ 93.3\pm1.4 $ & $ 92.1\pm0.9 $ & $ 94.3\pm1.2 $\\
$\ket{H}_\text{A}, \ket{V}_\text{B}$ & $ 94.5\pm1.2 $ & $ 95.2\pm1.4 $ & $ 92.1\pm1.2 $ & $ 94.6\pm1.8 $\\
\end{tabular}
\end{table}

In the $A_WB_WB_RA_R$ case, the read order of atoms $A$ and $B$ is swapped while all pulse timings are preserved. 
This results in storage times of $\unit[83]{\mu s}$ and $\unit[183]{\mu s}$ for atom $B$ and $A$, respectively.
For atom $B$, this leads to a similar performance. 
In contrast, atom $A$ shows a reduced fidelity for linear polarisation inputs, independent of the polarisation of atom $B$. 
We attribute this effect to magnetic-noise-induced decoherence which becomes relevant at this timescale as shown further below.

Regardless of the memory sequence, the observed fidelities are similar and largely independent of whether the input qubits are identical or orthogonal. 
This indicates a vanishing cross talk and therefore demonstrates the high degree of isolation of the two atoms. 
We remark that in a possible application where the whole sequence is retried until both memories successfully emitted a photon, the fidelity, as defined in Table \ref{table_fidelities}, increases on average by $\unit[0.6]{\%}$ (not shown). 

\begin{figure}
\includegraphics{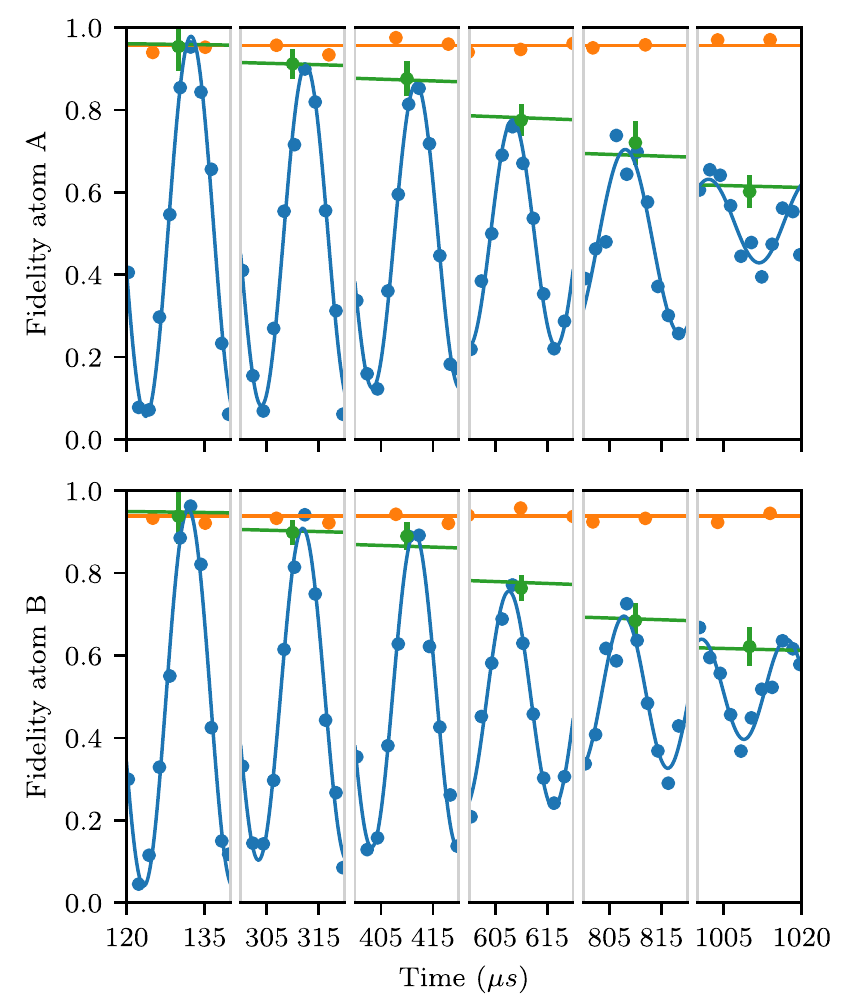}
\caption{
\label{fig_coherence} 
\textbf{Coherence time for both atoms.} Time evolution of the fidelity for the storage of circular (orange) and linear (blue) polarisations for atom A and atom B.
The sinusoidal fits to the $\unit[20]{\mu s}$ slices are fit with a Gaussian decay (green) giving a coherence time of more than $\unit[800]{\mu s}$ for both atoms.
The error bars represent the $\unit[95]{\%}$ confidence intervals of the statistical uncertainties.}
\end{figure}

One of the most important properties of a quantum memory is the storage time that we test using the $A_WB_WA_RB_R$ sequence. To this end, we vary the storage time between writing ($A_WB_W$) and reading ($A_RB_R$) of both atoms. The results for atom $A$ and atom $B$ are depicted in Fig. \ref{fig_coherence}. Both atoms exhibit a coherence time of more than $\unit[800]{\mu s}$ for linear polarisations and no measurable decay of the fidelity for circular polarisations. As in the single-atom case, the coherence time is limited by magnetic-field fluctuations on the few mG level. In addition, the coherence time is reduced by spatially sampling the locally varying magnetic field. Indeed, the margin on the atom positions is chosen in a way to have a reasonable data rate (see Supplement for further details). It is worth noting, that we expect no degradation of the coherence time for an even larger number of atoms as both atoms sample already a reasonably large region of the cavity mode. 

\begin{figure}
\includegraphics{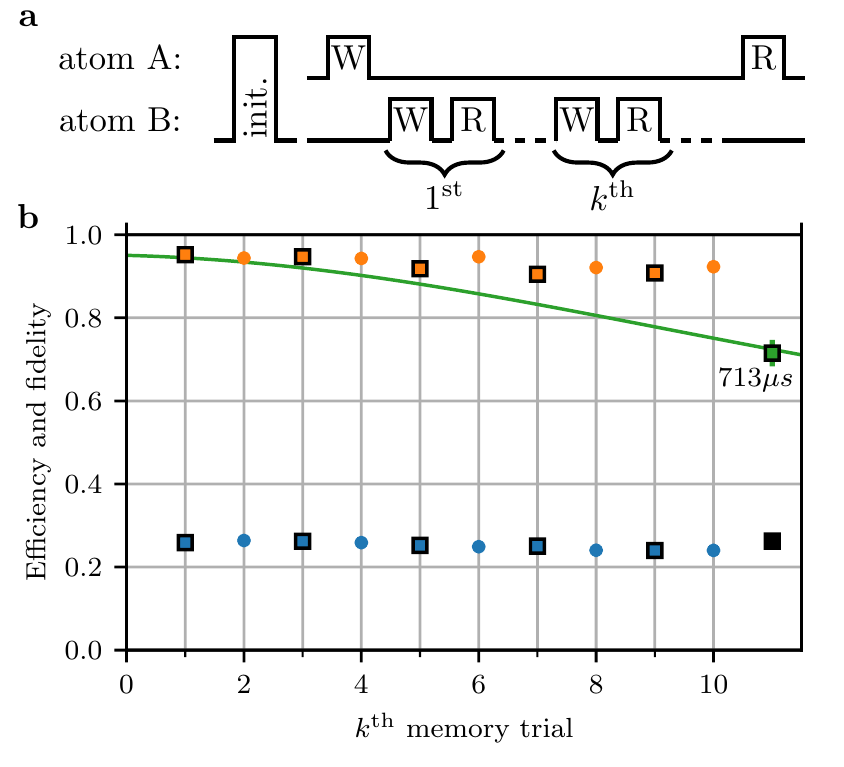}
\caption{
\textbf{Extended $\textbf{A}_\mathrm{W}\textbf{B}_\mathrm{W}\textbf{B}_\mathrm{R}\textbf{A}_\mathrm{R}$ scenario}.
\textbf{a}, 
After initializing both atoms, a qubit is written on atom A. 
During its storage time, different qubits are stored and retrieved on atom B. 
After 10 attempts (successful or not) on atom B, atom A's qubit is retrieved again. 
\textbf{b}, 
Fidelity and efficiency for qubits retrieved from atom A (green, black) and B (orange, blue), respectively.
Black circumferences denote linear input polarisations.
The green guide to the eye is taken from Fig. \ref{fig_coherence} to illustrate the decay in fidelity of atom A due to residual magnetic-field fluctuations. 
Linear fits to efficiency and fidelity give a decay per intermediate write-read process of $\unit[0.29]{pp}$ and $\unit[0.44]{pp}$, respectively.
The error bars represent the $\unit[95]{\%}$ confidence intervals of the statistical uncertainties.
}
\label{fig:reusing} 
\end{figure}

An extended case of the scenario $A_WB_WB_RA_R$ is presented in Fig. \ref{fig:reusing}. 
Here, atom $B$ is used multiple times during the storage of atom $A$.
This tests the cross talk for systems where many qubits are stored and retrieved serially on a single atom as shown here and also for future implementations where the number of atoms is scaled up even further.
Additionally it demonstrates that each atom can be used multiple times without re-initialization, allowing for higher repetition rates and potentially eliminating the need for atom-selective initialization.
In orange and blue, Figure \ref{fig:reusing}b shows the achieved fidelity and efficiency, respectively, for the reused atom $B$ as a function of number of trials.
Efficiency and fidelity decay by $\unit[0.29]{pp}$ and $\unit[0.44]{pp}$ per trial as predicted by theory (see Supplement).
Figure \ref{fig:reusing}b also shows the efficiency and fidelity of atom $A$ after the 10 memory trials on atom $B$.
Although only about $\unit[20]{\%}$ of the trials resulted in a successful memory event, in every trial the cavity is populated and the classical control field is applied, potentially leading to cross talk as discussed above.
Nevertheless, as the achieved fidelity is in good agreement with the expectation from the coherence-time measurement (green guide to the eye), any cross talk can be neglected even after 10 trials.\\

\textbf{OUTLOOK}\\

We have demonstrated an efficient network-ready random-access multi-qubit memory using two individually addressable single atoms.
We have identified and solved multiple challenges such as qubit cross talk which arise when making the most important step from one to several atoms in the same high-finesse optical cavity.
Given a lower bound on the store-and-retrieve efficiency of $\unit[20]{\%}$ and a cavity mode waist of $\unit[30]{\mu m}$, up to 5 atoms can be accommodated.
We estimate that this number can be easily increased to more than 20 atoms while still using the same trapping geometry, e.g. by using a tighter focus of the addressing beam while still using the same objective. As all of those atoms would still be coupled to the same cavity mode, cavity-mediated multi-qubit gates \cite{Borregaard2015,Welte2018} enable computation on all of them.
For future implementations, deterministic and thus zero-tolerance atom positioning, as can be achieved in optical tweezer arrays \cite{Endres2016,Barredo2018}, would further increase capacity, data rate and fidelity.
Also, it would allow for Zeeman selective coherent driving addressed onto single atoms, resulting in more than $\unit[100]{ms}$ coherence time \cite{Korber2018}.
In combination with atom-photon entanglement and atom-atom gates \cite{Reiserer2015,Welte2018}, the extension to multiple individually addressable atoms renders the cavity platform a promising candidate for future quantum networks with quantum repeaters \cite{Briegel1998} and distributed quantum computers \cite{Jiang2007}.\\

\textbf{DATA AVAILABILITY}
The data that support the findings of this study are available from the authors upon request.\\

\textbf{ACKNOWLEDGEMENTS}
We thank A. Neuzner and S. Ritter for contributions during the early stage of the experiments.
This work was supported by the Bundesministerium f\"{u}r Bildung und Forschung via the Verbund Q.Link.X (16KIS0870), by the Deutsche Forschungsgemeinschaft under Germany’s Excellence Strategy – EXC-2111 – 390814868, and by the European Union’s Horizon 2020 research and innovation programme via the project Quantum Internet Alliance (QIA, GA No. 820445).\\

\textbf{AUTHOR CONTRIBUTIONS}
All authors conceived the experiment. Experimental data were taken  by S.L., O.M. and M.K..
The manuscript was written by S.L., O.M. and G.R., with input from all authors.

\bibliographystyle{plain}

\let\oldaddcontentsline\addcontentsline
\renewcommand{\addcontentsline}[3]{}

\let\addcontentsline\oldaddcontentsline

\newpage
\onecolumngrid
\pagebreak

\topmargin 0.0cm
\oddsidemargin 0.2cm
\textwidth 17cm 
\textheight 21cm
\footskip 1.0cm

\renewcommand\thefigure{S\arabic{figure}} 
\renewcommand\refname{Supplementary References}
\renewcommand{\thesection}{S\arabic{section}}

\setcounter{equation}{0}
\setcounter{figure}{0}
\setcounter{table}{0}
\setcounter{page}{1}

\begin{center}
  \textbf{\large Supplemental Material: \\ A network-ready random-access qubits memory}\\[.2cm]
  S. Langenfeld, O. Morin,  M. Körber, and G. Rempe\\[.1cm]
  {\itshape Max-Planck-Institut f\"{u}r Quantenoptik, Hans-Kopfermann-Strasse 1, 85748 Garching, Germany\\}
\end{center}

\tableofcontents
\newpage

\begin{center}
\textbf{Supplementary Notes}
\end{center}

\section{Setup}
\subsection{Trapping and Sequence}
Atoms are loaded into a two-dimensional (2D) standing wave trap, consisting of a blue-detuned ($\lambda=\unit[772]{nm}$) intra-cavity trap (y-axis) and a red-detuned ($\lambda=\unit[1064]{nm}$) trap perpendicular to the cavity axis ($x$-axis). The trap frequencies are $2\pi\unit[(220,335,3.3)]{kHz}$ for the (x,y,z)-axis. After loading, there is a random number of atoms ($\bar{n}_\text{atom}\geq2$) with random positions inside the cavity.
The position of each atom is recorded via an imaging system consisting of an objective ($\text{NA}=0.43$) and an EMCCD camera \cite{S-NeuznerThesis}. 
A pair of atoms with distance between $\unit[6]{\mu m}$ and $\unit[20]{\mu m}$ is selected (the choice of the atom positions is explained in the main text and discussed in Section \ref{sec:acceptance_range}). All other atoms are heated out of the trap by applying resonant light only onto these atoms via the optical addressing system (see Section \ref{sec:addressing}). The selected pair is positioned symmetrically to the cavity center by shifting the red-detuned standing wave forming the trapping lattice and, along the blue-detuned intracavity trap axis, by (re)centering the atoms by decreasing and re-increasing the blue-detuned trap depth.
The typical $1/e$ trapping time for two atoms is $\unit[10]{s}$. Reloading of two atoms takes $\unit[2]{s}$ while only about 10\% of the reloading attempts give a usable atomic pattern.

As soon as a usable pair of atoms is trapped, we perform the experimental sequence consisting of pumping to the initial ground state $\ket{F=1, m_\mathrm{F}=0}$ including Raman resolved sideband cooling to the motional ground state along the cavity axis ($\unit[800]{\mu s}$), qubit storage and retrieval. Afterwards we perform polarization gradient cooling to produce fluorescence light for the atom imaging and to extend the atom trapping time.
Depending on the performed experiment, this sequence is repeated at a rate of $\unit[250-500]{Hz}$.

\subsection{Optical addressing system}
\label{sec:addressing}
\begin{figure}[hbt]
\includegraphics{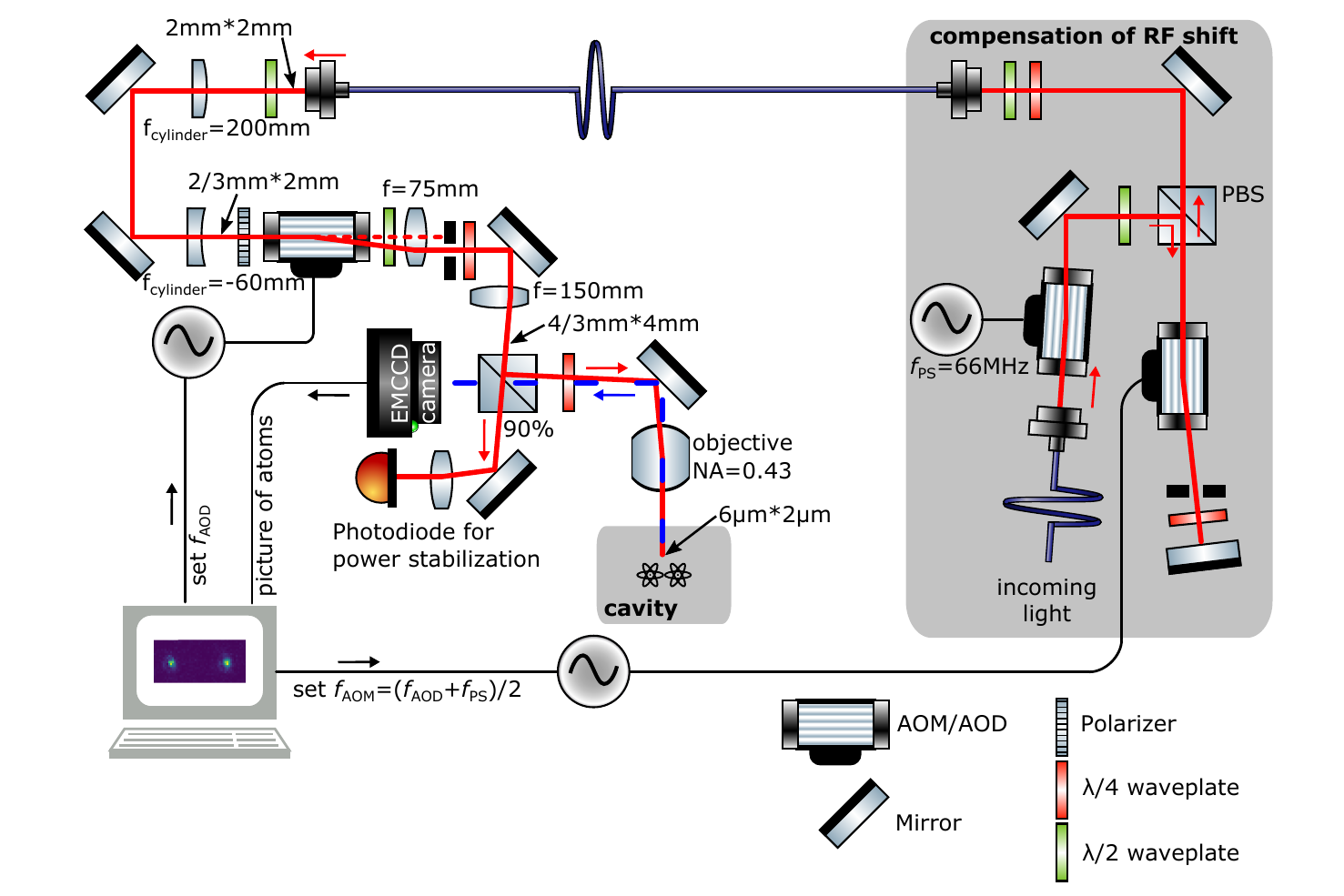}
\caption{
\label{fig_setup_detailed} 
\textbf{Detailed optical and electronic setup of the addressing system.}
The procedure in chronological order: After atoms have been loaded into the cavity, pictures of the atoms are evaluated via a computer/field-programmable-gate-array (FPGA) which subsequently sets the radio-frequency (RF) of the AOD so that it steers incoming light on the individual atoms. Additionally, the RF frequency of a preceding AOM is set, so that, in combination with another static-frequency AOM, it compensates the RF shift of the AOD. In this way, any incoming light retains its initial frequency.
The optical setup is tuned so that the light reaches the atoms with $\pi$-polarization and a focal spot of about $\unit[6]{\mu m}*\unit[2]{\mu m}$ (for the reason of asymmetry see Sec. \ref{sec:addressing_profile}).
}
\end{figure}

The optical and electronic setup is detailed in Fig. \ref{fig_setup_detailed}.
The high-NA objective used for imaging the atoms is also used for applying light fields selectively onto single atoms.
To this end, a non-polarizing beamsplitter (90:10) separates the track into imaging and addressing.
After recording the position of the atoms to be addressed, the radio-frequency (RF) input of an acousto-optic deflector (AOD) is tuned so that its deflection aims on a specific atom.
At the same time, the frequency of a preceding acousto-optical modulator (AOM) in a double-pass configuration is tuned so that it compensates the varying RF shift due to the AOD.

\newpage

\section{Cross talk via the common cavity mode}
\begin{figure}[hbt]
\begin{tikzpicture}[scale=1]
		\node[inner sep=0pt] (cavity) at (0,0) {\includegraphics{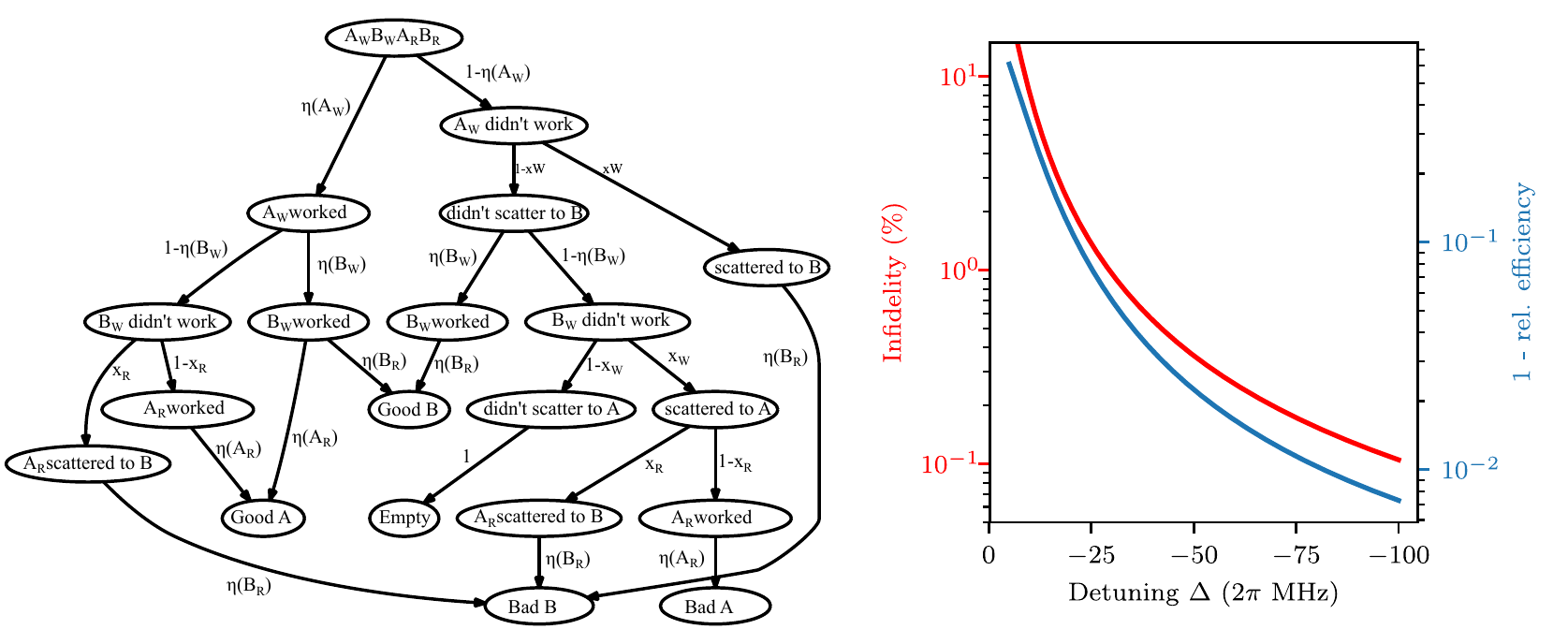}};
		\definecolor{lightblue}{RGB}{31,119,180};
		\definecolor{red}{RGB}{255,0,0};
		\node at (-8.2,3.1) {\textbf{a}};
		\node at (0,3.1) {\textbf{b}};
\end{tikzpicture}
\caption{
\label{fig_gorshkov_detuning} 
\textbf{Cross talk via the common cavity mode.}
\textbf{a},
Tree of different possible outcomes of $\aw\bw\ar\br$ memory protocol, depending on the detuning-dependent write-in efficiency $\eta(\aw,\bw)$ and read-out efficiency $\eta(\ar,\br)$. Here $x_\mathrm{R}$ and $x_\mathrm{W}$ are the detuning-dependent scattering probabilities to the unaddressed atom for read-out and write-in, respectively. 
All parameters are simulated \cite{S-Morin2019} and verified experimentally for the detunings given in the main text.
\textbf{b}, 
Infidelity and reduction in efficiency due to this scattering versus single-photon detuning per photon entering the cavity. 
At small detunings, this scattering renders a qubit memory impossible.
For detunings larger than $\unit[100]{MHz}$, the infidelity becomes negligible for our application. Arrows indicate to which of the y-axes the curves correspond.
}
\end{figure}

As described in the main text and also the caption of Fig. 1 of the main text, one cross talk mechanism arises from the fact that both atoms are identically coupled to the same cavity field.

In one scenario, both atoms are initialized in $\is$ and thus are ready to accept a photon. As described in the main text, even the not-addressed atom can interact with the cavity population via off-resonant scattering.
Figure \ref{fig_gorshkov_detuning}a depicts a tree of possible outcomes of an $\aw\bw\ar\br$ memory. At the different stages, either the write/read worked with probability $\eta(\aw,\bw),\eta(\ar,\br)$, or if it didn't work, there is a certain probability $x_\mathrm{R},x_\mathrm{W}$ that the photon scattered to the additional (not-addressed-) atom.
Figure \ref{fig_gorshkov_detuning}b shows the derived fidelity and efficiency as a function of single photon detuning $\Delta$. The observed scaling is only possible because we were able to make $\eta(\aw,\bw),\eta(\ar,\br)$ almost detuning independent \cite{S-Morin2019}.
Thus, for all presented measurements, a detuning of $\Delta/2\pi=\unit[-100]{MHz}$, limited by available laser power, was chosen.

In a second scenario, the second, not-addressed, atom already successfully stores a qubit.
Due to the large detuning of $\Delta=\Delta_\text{hf}\approx \unit[6.8]{GHz}$ between the initial and the storage state, neither scattering nor ac-Stark shifts on the stored qubit contribute significantly.

\section{Cross-illumination}

\begin{figure}[H]
\begin{tikzpicture}[scale=1]
    \node[inner sep=0pt] (cavity) at (0,0) {\includegraphics{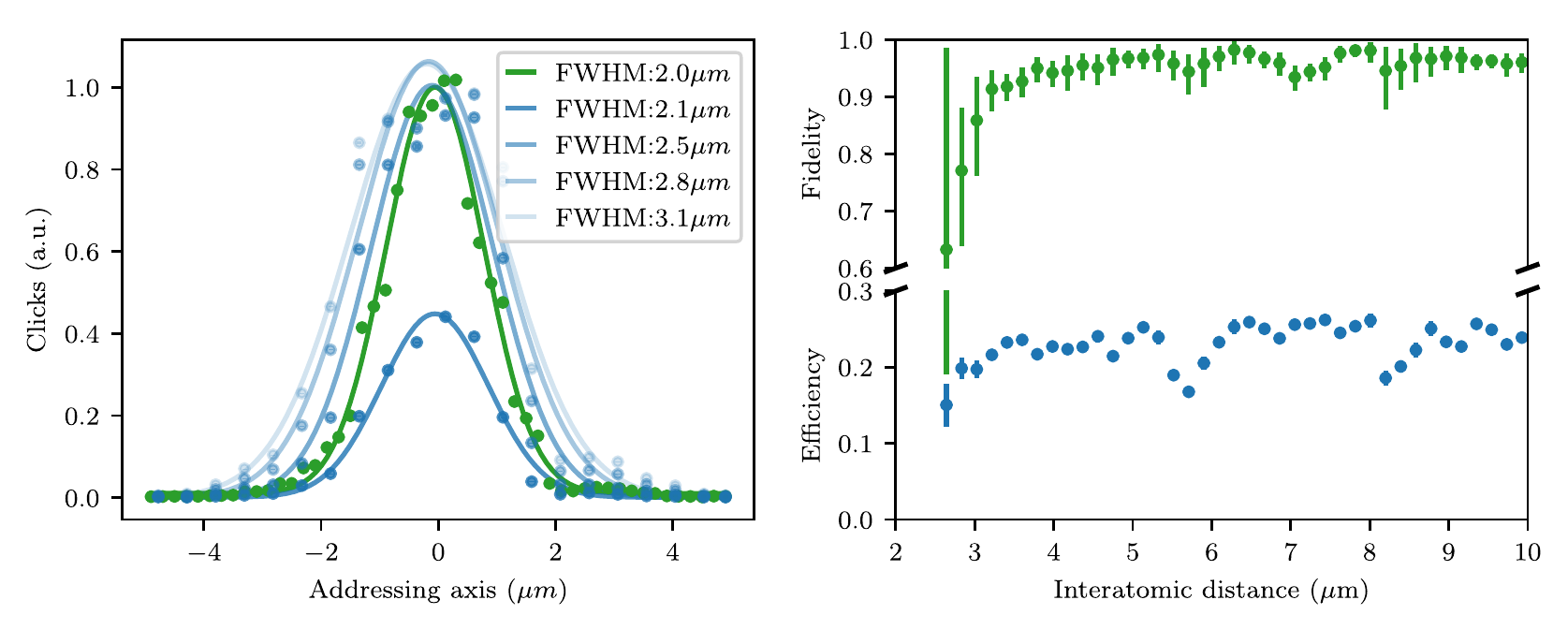}};
		\node at (-8.2,3.1) {\textbf{a}};
		\node at (0,3.1) {\textbf{b}};
\end{tikzpicture}
\caption{
\label{fig_profile} 
\textbf{Optical profile of addressing system.}
\textbf{a}, 
The green curve shows the intensity profile of the addressing beam. The blue curves show how the efficiency and effective beam width of the photon production change when STIRAP control pulses of different lengths are applied via the addressing system.
\textbf{b},
Fidelity and efficiency of the first atom in an $\aw\bw\ar\br$ protocol where opposite circular polarizations are stored on atom A and B.
For distances on the order of the addressing beam width, cross-illumination of the two atoms leads to a decrease in fidelity.
The efficiency is less affected and stays about constant within systematic error over the investigated region.
}
\end{figure}

\subsection{Addressing beam profile}
\label{sec:addressing_profile}
The addressing beam is used to apply light fields onto atoms which are distributed along the $x$-axis and centered along the cavity- ($y$) axis.
To this end, the width of this beam along the addressing axis should be significantly smaller than the expected atom-atom distance, which itself is upper limited by the cavity mode size of $\unit[30]{\mu m}$ \cite{S-NeuznerThesis}.
However, in order to be less susceptible to mechanical drifts of the setup, it is beneficial to have the addressing waist bigger than its theoretical minimum of $\unit[0.6]{\mu m}$.  
Here, we tuned the optical system to have a full width at half maximum (FWHM) in intensity of $\unit[(1.96\pm 0.04)]{\mu m}$.
As there is no addressing along the cavity ($y$) axis, the beam is elongated three-fold along this direction.
This allows to accept a larger range of possible atom positions resulting in higher data rates.
For further scaling to more atoms, there is no technological or physical reason not to have a narrower beam width in both dimensions.

\subsubsection{Intensity profile}
In order to verify the intensity profile of the addressing beam as seen by the atom, we aim the addressing beam at a fixed position and probe the local intensity of the light by varying the atom position within.
We apply resonant light on the cycling transition $\ket{F=2}\leftrightarrow\ket{F'=3}$ while having the cavity resonant with the same transition.
The number of scattered photons picked up by the cavity and detected on a photon counter is linear in intensity for intensities small compared to the saturation intensity \cite{S-Bochmann2010}.
In green, Fig. \ref{fig_profile}a shows the number of clicks in arbitrary units as a function of a single atom's position relative to the center of the addressing beam.
As given above, the FWHM is $\unit[(1.96\pm 0.04)]{\mu m}$.
The center of the beam is at the aimed-at position, due to the calibration of atom-detection EMCCD-pixel and AOD RF-frequency.

\subsubsection{STIRAP profile}
In the main text, only STIRAP control pulses are applied through the addressing system. 
As this is a non-linear process \cite{S-Gorshkov}, the relevant addressing profile differs from the intensity profile.
Here, we probe the profile again by aiming at a fixed position and changing the position of a single atom.
In contrast to above, the cavity frequency is chosen in the same way as for the main text, i.e. resonant close to the $\ket{F=1}\leftrightarrow\ket{F'=1}$ transition.
We globally prepare the storage state manifold $\ket{F=2}$ and afterwards apply a read-out STIRAP pulse through the addressing system.

The amplitude and shape of the emitted photon depend on the temporal intensity profile of the applied STIRAP control pulse   \cite{S-Gorshkov}.
Thus, the read-out control pulse has to be optimized to not saturate the atom but still read most of the population.
Figure \ref{fig_profile}a depicts the addressing profile for different read-out durations in different shades of blue.
For a duration of $\unit[7]{\mu s}$ (darkest blue), even a perfectly addressed atom cannot be read completely.
In contrast, a $\unit[12]{\mu s}$ (lightest blue) read-out duration gives a broadened, non-Gaussian profile of full amplitude.
For the experiments demonstrated in the main text and this supplement, a read-out duration of $\unit[8]{\mu s}$ (second darkest blue) was chosen, which gives a FWHM of $\unit[(2.5\pm0.1)]{\mu m}$ while preserving almost completely the process efficiency.

\subsection{Minimum interatomic distance}
Due to the finite size of the addressing beam, there has to be a minimum distance between the atoms so that cross-illumination is minimized.
On the other side, the distance should be as small as possible to either allow for a larger acceptance range for possible positions and thus higher data rate, or in future implementations for a greater number of atoms in a volume limited by the cavity mode size.
In order to get an application relevant bound for the interatomic distance, we perform a two-atom memory in the $\aw\bw\ar\br$ protocol.
Here, opposite circular polarisations are stored on atom A and B to not be affected by the decoherence of linear polarisations and to have maximum infidelity in case of cross-illumination.
Figure \ref{fig_profile}b shows the fidelity and efficiency of the read-out photons of atom A.
For distances below about $\unit[2.5]{\mu m}$, the imaging system cannot resolve two separated atoms, so that there is no addressing possible for these configurations.
Although this results in large error bars, the fidelity decreases significantly for interatomic distances below $\unit[3]{\mu m}$, in agreement with the addressing profiles presented above.
For distances above $\unit[5]{\mu m}$, the fidelity approaches single-atom fidelities \cite{S-Korber2018}.
In contrast, the efficiency shows no clear trend and stays about constant within the investigated region.
Due to these findings, all presented data is taken with a minimum distance of $\unit[6]{\mu m}$.

\section{Decoherence due to acceptance range of atomic positions}
\label{sec:acceptance_range}

\begin{figure}[H]
\begin{tikzpicture}[scale=1]
		
		\node[inner sep=0pt] (cavity) at (0,0) {\includegraphics{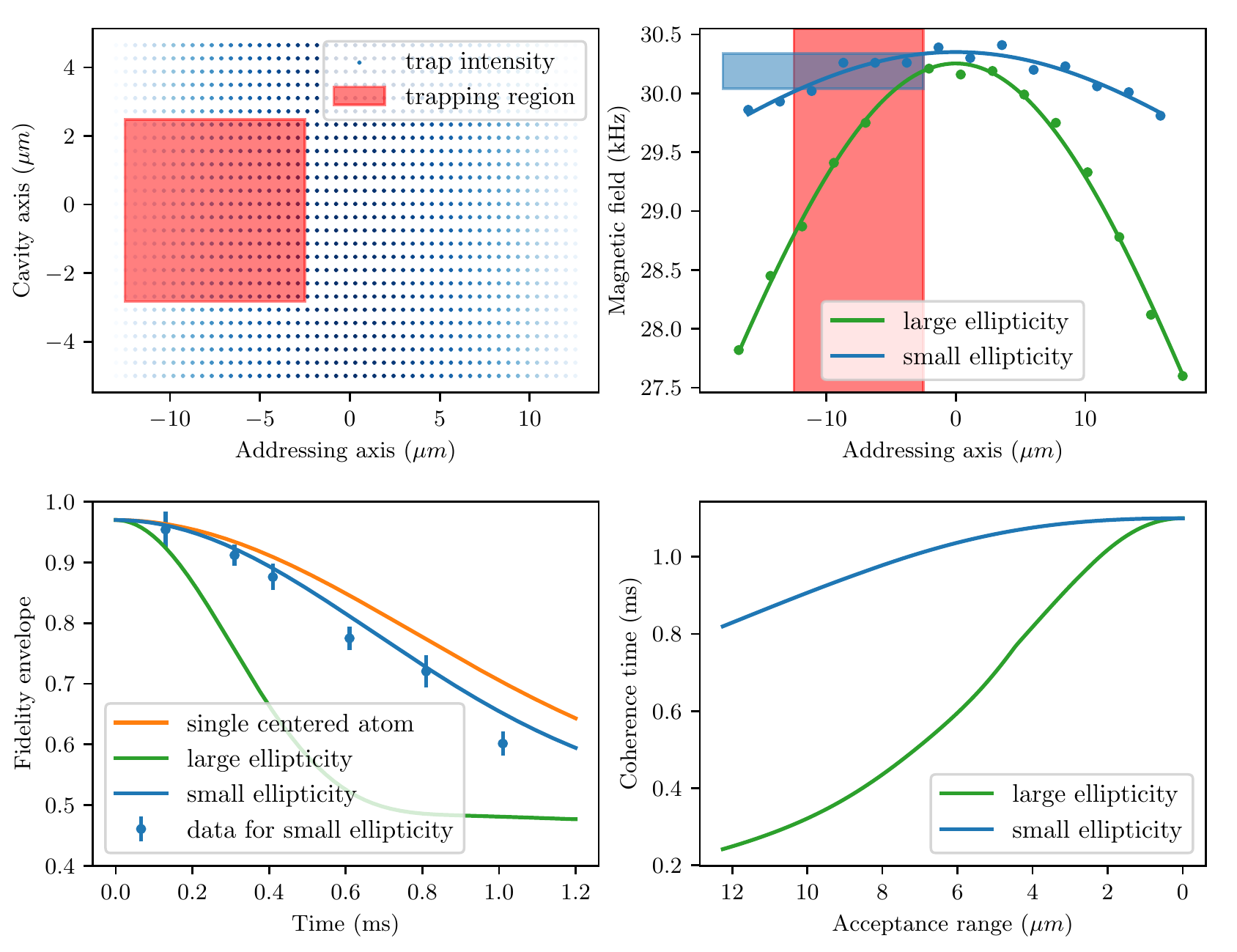}};
		\node at (-8.2,6.2) {\textbf{a}};
		\node at (0,6.2) {\textbf{b}};
		\node at (-8.2,-0.3) {\textbf{c}};
		\node at (0.0,-0.3) {\textbf{d}};
\end{tikzpicture}

\caption{
\label{fig_magneticfield} 
\textbf{Decoherence mechanism due to acceptance range of atomic positions. a}, 
The intra-cavity trap power is presented as a function of atom position in the cavity. The red-highlighted region shows the acceptance range for the data presented in the main text. 
\textbf{b}, 
Due to circular components of the trap light, atoms see different virtual magnetic fields as a function of position. Here we test this with two different arbitrary settings of ellipticity. The main text uses the small ellipticity setting.
\textbf{c}, 
This leads to dephasing of the memory, here depicted for different ellipticities and compared to the experimental data presented in the main text. 
\textbf{d}, 
Independent of the circularity, this decoherence can be minimized by fixing the atom positions, e.g. by using an optical tweezer.
}
\end{figure}

In the present scheme, atoms are loaded probabilistically into the two-dimensional trap inside the cavity.
For single-atom protocols, the atom can be positioned within this trap to an accuracy of about $\unit[1]{\mu m}$.
However, currently this positioning is limited to shifting all atoms in the trap simultaneously.
For multi-atom protocols, this results in a shiftable center-of-mass, whereas the relative position of the atoms is given by the initial loading or subsequent jumps between lattice sites.
In order to increase the data rate, a certain range of possible intermediate distances is accepted.
As illustrated in Fig. \ref{fig_magneticfield}a, this leads to a sampling of the intensity of the intra-cavity (blue-detuned) $\unit[772]{nm}$ trap.

Due to circular components of the intra-cavity trap light, the atoms see a virtual magnetic field that depends on the intensity\cite{S-NeuznerThesis}.
Figure \ref{fig_magneticfield}b shows the relevant magnetic field as seen by the atom for a given position on the addressing axis.
Depicted are two curves for two different ellipticities of the intra-cavity trap light.
As the atoms sample a certain region and thus intensity, they are subject to a varying magnetic field environment, leading to a range of different precession Larmor frequencies, and thus to dephasing.
In Fig. \ref{fig_magneticfield}c, the theoretical decay in fidelity over time is presented for the two different mentioned ellipticities, assuming equal sampling of the acceptance range used for the experiments demonstrated in the main text.
In the experiments presented in the main text, the situation labeled \textit{small ellipticity} is used.
For this configuration, Fig. \ref{fig_magneticfield}c also shows the experimental result of the decay in fidelity as presented in the main text.
Additionally, Fig. \ref{fig_magneticfield}d shows the theoretical coherence time as a function of acceptance range of atomic positions.
This clearly shows, that this is no fundamental limitation to the coherence time, as in principle the position of every atom could be fixed by e.g. employing an optical tweezer technique.
Alternatively, the oscillation frequency could be feedback by the known position of the atoms or, as mentioned in the main text, the oscillation could be completely compensated by a rotating waveplate/electro-optical modulator.

\newpage

\section{Extended $\aw\bw\br\ar$ scenario theory}

\begin{figure}[H]
\begin{tikzpicture}[scale=1]
		\node[inner sep=0pt] (cavity) at (0,0) {\includegraphics{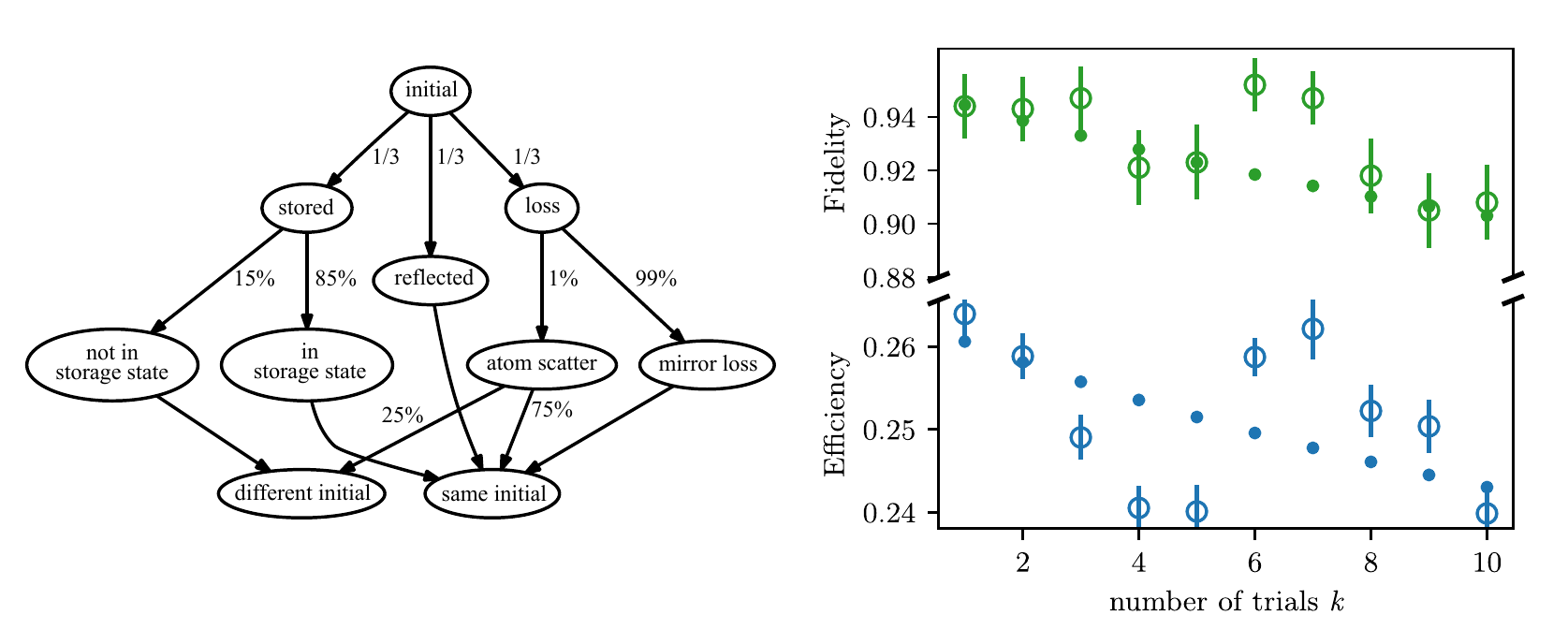}};
		\node at (-8.2,3.1) {\textbf{a}};
		\node at (0,3.1) {\textbf{b}};
\end{tikzpicture}
\caption{
\label{fig_reusingtheory} 
\textbf{Extended $\aw\bw\br\ar$ scenario theory.}
\textbf{a}, 
Markov process of the memory to test whether the atom ends up in the same initial state it started in, or goes to a different initial state within the $\ket{F=1}$ manifold after a write-read process.
The probability to stay in the same initial state is about $\unit[95]{\%}$.
\textbf{b}, 
Filled dots show the prediction of a model of the fidelity and efficiency for successive memory trials on the same atom without re-initialization for the storage and retrieval of circular polarization qubits. 
Open symbols denote experimental data taken from the main text.
All necessary parameters are extracted from independent measurements.
}
\end{figure}

Ideally, starting from the initial state $\is$, every write-read cycle ends again in the initial state.
If this would be the case, every repeated write-read attempt would result in the same efficiency and fidelity.
However, there are multiple experimental imperfections and physical side-effects which prevent this behavior.

We probe the memory with weak coherent pulses of average photon number $\bar{n}=1$.
Due to losses of the cavity, about $\unit[33]{\%}$ of the pulse do not interact with the atoms.
At a single-photon detuning of $\Delta/2\pi=\unit[-100]{MHz}$, the scattering of the atom contributes by about $\unit[0.3]{\%}$.
The branching ratio of the excited states leads to a decay to the initial state in $\unit[75]{\%}$ of the cases.
The remaining $\unit[66]{\%}$ of the incoming photon are evenly shared by reflection of the pulse and successful storage in the desired atom.

The main shortcoming is the non-zero incoherent scattering probability of every STIRAP process due to the finite atom-light interaction rate $g$.
For every read-out attempt of a successful write-in, the finite $g$ results in a reduced probability of $\unit[85]{\%}$ to go back to $\is$, whereas the remaining $\unit[15]{\%}$ are distributed over $\ket{F=1, m_\mathrm{F}=\pm1}$, depending on the population of the storage states.
Note that these decay processes do not contribute to the reported single-trial performance, as the cavity does not support the read-out photons of this polarization.
Note that repeated read-out does not increase the efficiency, as the final state of every read-out attempt is in the $F=1$ manifold.
Starting from one of the states $\ket{F=1, m_\mathrm{F}=\pm1}$, the next memory only performs with an efficiency of $\unit[14]{\%}$ and an initial fidelity of $\unit[76]{\%}$, again depending on a combination of starting state and input polarization.
This memory-trial again has some probability to end up in either the desired initial state $\is$, or in one of $\ket{F=1, m_\mathrm{F}=\pm1}$.
Assuming the same probabilities as mentioned above, i.e. $\unit[85]{\%}$ for returning to the initial state of this memory attempt, we set up a Markovian chain of events.
Figure \ref{fig_reusingtheory}a shows a schematic summary of the described processes.
Figure \ref{fig_reusingtheory}b presents the evolution of fidelity and efficiency versus the number of write-read attempts as experimentally demonstrated in the main text. There is a reasonable agreement between theory and experiment.

\bibliographystyle{plain}

\begin{thebibliography}{10}

\bibitem{Wootters1982} Wootters, W. K. and Zurek, W. H. {A single quantum cannot be cloned}. \textit{Nature} \textbf{299}, 802--803 (1982).

\bibitem{Zhang2018} Zhang, Q., Xu, F., Chen, Y.-A., Peng, C.-Z., and Pan, J.-W. {Large scale quantum key distribution: challenges and solutions}. \textit{Opt. Express} \textbf{26}, 24260--24273 (2018).

\bibitem{Yin2020} Yin, J. \textit{et al.} {Entanglement-based secure quantum cryptography over 1,120 kilometres}. \textit{Nature} \textbf{582}, 501--505 (2020).

\bibitem{Kimble2008} Kimble, H. J. {The quantum internet}. \textit{Nature} \textbf{453}, 1023--1030 (2008).

\bibitem{Wehner2018} Wehner, S., Elkouss, D. and Hanson, R. {Quantum internet: A vision for the road ahead}. \textit{Science} \textbf{362}, eaam9288 (2018).

\bibitem{VanEnk1997} van Enk, S. J., Cirac, J. I. and Zoller, P. {Ideal Quantum Communication over Noisy Channels: A Quantum Optical Implementation}. \textit{Phys. Rev. Lett.} \textbf{78}, 4293 (1997).

\bibitem{Briegel1998} Briegel, H.-J., D{\"{u}}r, W., Cirac, J. I. and Zoller, P. {Quantum Repeaters: The Role of Imperfect Local Operations in Quantum Communication}. \textit{Phys. Rev. Lett.} \textbf{81}, 5932 (1998).

\bibitem{Childress2006} Childress, L., Taylor, J. M., S{\o}rensen, A. S. and Lukin, M. D. {Fault-tolerant quantum communication based on solid-state photon emitters}. \textit{Phys. Rev. Lett.} \textbf{96}, 070504 (2006).

\bibitem{Jiang2007} Jiang, L., Taylor, J. M., S{\o}rensen, A. S. and Lukin, M. D. {Distributed quantum computation based on small quantum registers}. \textit{Phys. Rev. A} \textbf{76}, 1--22 (2007).

\bibitem{DiVincenzo2000} DiVincenzo, D. P. {The Physical Implementation of Quantum Computation} \textit{Fortschr. Phys.} \textbf{48}, 9-11 (2000).

\bibitem{Landauer1995} Landauer, R. {Is Quantum Mechanics Useful?}  \textit{Philos. Trans. Royal Soc. A} \textbf{353}, 367–376 (1995).

\bibitem{Sinclair2014} Sinclair, N. \textit{et al.} {Spectral Multiplexing for Scalable Quantum Photonics using an Atomic Frequency
Comb Quantum Memory and Feed-Forward Control}. \textit{Phys. Rev. Lett.} \textbf{113}, 053603 (2014).

\bibitem{Tang2015} Tang, J.-S. \textit{et al.} {Storage of multiple single-photon pulses emitted from a quantum dot in a solid-state quantum memory}. \textit{Nat. Commun.} \textbf{6}, 8652 (2015).

\bibitem{Laplane2015} Laplane, C. \textit{et al.} {Multiplexed on-demand storage of polarization qubits in a crystal} \textit{New J. Phys.} \textbf{18}, 013006 (2016).

\bibitem{Tiranov2016} Tiranov, A. \textit{et al.} {Temporal Multimode Storage of Entangled Photon Pairs}. \textit{Phys. Rev. Lett.} \textbf{117}, 240506 (2016).

\bibitem{Jiang2019} Jiang, N. \textit{et al.} {Experimental realization of 105-qubit random access quantum memory}. \textit{Npj Quantum Inf.} \textbf{5}, 1 (2019).

\bibitem{Li2020} Li, C. \textit{et al.} {Quantum Communication between Multiplexed Atomic Quantum Memories}. \textit{Phys. Rev. Lett.} \textbf{124}, 240504 (2020).

\bibitem{Saffman2016} Saffman, M., {Quantum computing with atomic qubits and Rydberg interactions: progress and challenges}. \textit{J. Phys. B: At. Mol. Opt. Phys.} \textbf{49}, 202001 (2016).

\bibitem{Reiserer2015} Reiserer, A. and Rempe, G. {Cavity-based quantum networks with single atoms and optical photons}. \textit{Rev. Mod. Phys.} \textbf{87}, 1379--1418 (2015).

\bibitem{Korber2018} K{\"{o}}rber, M. \textit{et al.} {Decoherence-protected memory for a single-photon qubit}. \textit{Nat. Photonics} \textbf{12}, 18--21 (2018).

\bibitem{Wilk2010} Wilk, T. \textit{et al.} {Entanglement of Two Individual Neutral Atoms Using Rydberg Blockade}. \textit{Phys. Rev. Lett.} \textbf{104}, 010502 (2010).

\bibitem{Kaufman2015} Kaufman, A. M. \textit{et al.} Entangling two transportable neutral atoms via local spin exchange. \textit{Nature} \textbf{527}, 208--211 (2015).

\bibitem{Weiss2017} Weiss, D. S. and Saffman, M. {Quantum computing with neutral atoms}. \textit{Phys. Today} \textbf{70}, 44--50 (2017).

\bibitem{Levine2019} Levine, H. \textit{et al.} {Parallel Implementation of High-Fidelity Multiqubit Gates with Neutral Atoms}. \textit{Phys. Rev. Lett.} \textbf{123}, 170503 (2019).

\bibitem{Bruzewicz2019} Bruzewicz, C. D., Chiaverini, J., McConnell, R. and Sage, J. M. {Trapped-ion quantum computing: Progress and challenges}. \textit{Appl. Phys. Rev.} \textbf{6}, 021314 (2019).

\bibitem{Casabone2015} Casabone, B. \textit{et al.} {Enhanced Quantum Interface with Collective Ion-Cavity Coupling}. \textit{Phys. Rev. Lett.} \textbf{114}, 023602 (2015).

\bibitem{Inlek2017} Inlek, I. V., Crocker, C., Lichtman, M., Sosnova, K. and Monroe, C. {Multi-Species Trapped Ion Node for Quantum Networking.} \textit{Phys. Rev. Lett.} \textbf{118}, 250502 (2017). 

\bibitem{Welte2018} Welte, S., Hacker, B., Daiss, S., Ritter, S. and Rempe, G. {Photon-Mediated Quantum Gate between Two Neutral Atoms in an Optical Cavity}. \textit{Phys. Rev. X} \textbf{8}, 011018 (2018).

\bibitem{Nunn2008} Nunn, J. \textit{et al.} {Multimode Memories in Atomic Ensembles}. \textit{Phys. Rev. Lett.} \textbf{101}, 260502 (2008).

\bibitem{Neuzner2016} Neuzner, A., K{\"{o}}rber, M., Morin, O., Ritter, S. and Rempe, G. {Interference and dynamics of light from a distance-controlled atom pair in an optical cavity}. \textit{Nat. Photonics} \textbf{10}, 303--306 (2016).

\bibitem{Morin2019} Morin, O., K{\"{o}}rber, M., Langenfeld, S. and Rempe, G. {Deterministic Shaping and Reshaping of Single-Photon Temporal Wave Functions}. \textit{Phys. Rev. Lett.} \textbf{123}, 133602 (2019).

\bibitem{Borregaard2015} Borregaard J., Kómár P., Kessler E. M., Sørensen A. S., and Lukin, M. D. {Heralded Quantum Gates with Integrated Error Detection in Optical Cavities}. \textit{Phys. Rev. Lett.} \textbf{114}, 110502 (2015).

\bibitem{Endres2016} Endres, M. \textit{et al.} {Atom-by-atom assembly of defect-free one-dimensional cold atom arrays}. \textit{Science} \textbf{354}, 1024--1027 (2016).

\bibitem{Barredo2018} Barredo, D., Lienhard, V., de L{\'{e}}s{\'{e}}leuc, S., Lahaye, T. and Browaeys, A. {Synthetic three-dimensional atomic structures assembled atom by atom}. \textit{Nature} \textbf{561}, 79--82 (2018).
\end{thebibliography}

\begin{thebibliography}{}

\bibitem{S-NeuznerThesis} Neuzner, A. \textit{Resonance Fluorescence of an Atom Pair in an Optical Resonator}, PhD Thesis, TU M{\"{u}}nchen (2015).

\bibitem{S-Morin2019} Morin, O., K{\"{o}}rber, M., Langenfeld, S. and Rempe, G. {Deterministic Shaping and Reshaping of Single-Photon Temporal Wave Functions}.
\textit{Phys. Rev. Lett.} \textbf{123}, 133602 (2019).

\bibitem{S-Bochmann2010} Bochmann, J. \textit{et al.} {Lossless State Detection of Single Neutral Atoms.}
\textit{Phys. Rev. Lett.} \textbf{104}, 203601 (2010).

\bibitem{S-Gorshkov} Gorshkov, A. V., Andre, A., Lukin, M. D. and S{\o}rensen, A. S. {Photon storage in $\Lambda$-type optically dense atomic media. I. Cavity model.}
\textit{Phys. Rev. A} \textbf{76}, 033804 (2007).

\bibitem{S-Korber2018} K{\"{o}}rber, M. \textit{et al.} {Decoherence-protected memory for a single-photon qubit}.
\textit{Nat. Photonics} \textbf{12}, 18--21 (2018).

\end{thebibliography}

\end{document}